\documentstyle[preprint,aps,epsf]{revtex}

\newcommand{\ewxy}[2]{\setlength{\epsfxsize}{#2}\epsfbox[10 30 640
  590]{#1}}

\begin{document}

\draft

\vspace{2cm}

\title{ON THE ROLE OF THE EFFECTIVE INTERACTION IN
QUASI--ELASTIC ELECTRON SCATTERING CALCULATIONS}

\author{Eduardo Bauer \cite{AAAuth}}

\address{Departamento de F\'{\i}sica, Facultad de Ciencias
Exactas, Universidad Nacional de La Plata, \\La Plata, 1900,
Argentina.}

\author{Antonio M. Lallena}

\address{Departamento de F\'{\i}sica Moderna, Universidad de
Granada, E--18071 Granada, Spain.}

\maketitle

\begin{abstract}
The role played by the effective residual interaction in the
transverse nuclear response for quasi--free electron scattering
is discussed. The analysis is done by comparing different calculations
performed in the Random--Phase Approximation and Ring Approximation 
frameworks. The importance of the exchange terms in this energy region
is investigated and the changes on the nuclear responses due to the
modification of the interaction are evaluated.
The calculated quasi--elastic responses show clear indication of their
sensibility to the details of the interaction and this imposes the
necessity of a more careful study of the role of the different
channels of the interaction in this excitation region. 
\end{abstract}

\vspace{.5cm}


\narrowtext

\section{INTRODUCTION}

An aspect of great importance in nuclear structure calculations at any
excitation energy concerns with the role of the effective interaction.
At low energies this problem has generated a considerable body of work
in the last twenty years. On the contrary, this is a question not 
studied yet in deep in the literature for higher energies.

Giant resonances show an intricate mixture of multipolarities and the
study of how the interaction affects it is a difficult
task. In the quasi--elastic peak region, the problem of the
longitudinal and transverse separation has occupied most of the
investigations carried out till now and the discussion of the effects
due to changes in the effective interaction have not been considered
in detail.

As an example, we mention the considerable number of Random--Phase 
Approximation (RPA) type calculations performed in this energy region,
much of them using residual interactions which include basically a 
zero--range term plus meson--exchange potentials corresponding to
$\pi$, $\rho$ and, eventually, other mesons~\cite{al84}--\cite{gi97}. 
An important point concerning the interaction refers to the values
chosen for the parameters entering in the zero--range piece.
However, and to the best of our knowledge, only in
Ref.~\cite{gi97} a certain discussion relative to the effects of
varying these parameters can be found. In fact, the common
practice is to pick an interaction from the literature, which
usually corresponds to a parameterization fixed for low energy
calculations, and afterwards use it to evaluate quasi--free
observables sometimes without taking care of the effective theory in 
which the interaction was adjusted. It is obvious that, to a certain
level, doubtful results are possible because of the known link
between effective theory and interaction.

In this work we want to address this question and investigate if 
different parametrizations of the interaction can produce noticeable 
differences in the results and the extent to which the use of an 
interaction fixed for a given effective theory affects the results 
obtained within a different one. In Sec.~II we give the details about 
the effective theories and interactions used to perform the
calculations. In Sec.~III we show and discuss the results we have
obtained. Finally, we present our conclusions in Sec.~IV.

\section{EFFECTIVE THEORIES AND INTERACTIONS}

The first interaction we consider in this work is the so--called 
J\"ulich--Stony Brook interaction~\cite{sp80} which is an effective 
force widely used for calculations in the quasi--elastic peak. It is 
given as follows: 
\begin{equation}
\label{int-I}
\displaystyle
V^{\rm I}_{\rm res} \, = \, V_{\rm LM} \, + \, V_\pi \, + \,
\tilde{V}_\rho \, . 
\end{equation}
Here $V_{\rm LM}$ is a zero--range force of Landau--Migdal type,
which takes care of the short--range piece of the NN
interaction: %
\begin{equation}
V_{\rm LM} \, = \,C_0 \,[ g_0\, 
{\mbox{\boldmath $\sigma$}}^1\cdot {\mbox{\boldmath $\sigma$}}^2 
+\, g_0^\prime\,
{\mbox{\boldmath $\sigma$}}^1\cdot {\mbox{\boldmath $\sigma$}}^2
{\mbox{\boldmath $\tau$}}^1  \cdot {\mbox{\boldmath $\tau$}}^2]
\, .
\end{equation}
On the other hand, a finite--range component generated by the
($\pi$+ $\rho$)--meson exchange potentials is also included. The
tilde in $\tilde{V}_\rho$ means that the bare $\rho$--exchange
potential is slightly modified in order to take into account the
effect of the exchange of more massive mesons. In particular, a 
factor $r=0.4$ is multiplying the finite--range non--tensor piece
of the potential (see Ref.~\cite{sp80} for details). This force
was fitted to reproduce low energy magnetic properties in the
lead region (specifically, magnetic resonances in $^{208}$Pb and
magnetic moments and transition probabilities in the
neighboring nuclei). The calculations were performed in the
framework of the RPA and Woods--Saxon single--particle wave
functions were used in the configuration space. The values 
$g_0=0.57$ and $g_0^\prime=0.717$ (with $C_0=386.04$~MeV~fm$^3$) 
were found to be adequate to describe the properties 
studied.

As previously stated, this interaction has been considered in 
different calculations in the quasi--elastic peak (see
e.g.~\cite{gi97}). The problem arise because some of them have been
done within the Fermi gas (FG) formalism, with local density
approximation to describe finite nuclei, in the ring approximation 
(RA), where the exchange terms are not taken into account, and with 
the full unmodified $\rho-$exchange potential. Under these 
circumstances, the possible effects in the nuclear 
responses due to the inconsistency between the model and the effective
interaction could be non--negligible. This is precisely the first aspect
we want to investigate. To do that we compare the responses obtained
with the J\"ulich--Stony Brook interaction with those calculated with
a second effective force of the form: 
\begin{equation}
\label{int-II}
\displaystyle
V^{\rm II}_{\rm res} \, = \, V_{\rm LM} \, + \, V_\pi \, + \,
V_\rho \, ,
\end{equation}
by considering the same values for the zero--range parameters in both
cases. The force in Eq.~(\ref{int-II}) only differs from 
$V^{\rm I}_{\rm res}$ in the $\rho$-potential which, in this case,
does not include any reduction factor. Both RPA and RA effective
theories are used to analyze the results.

A second question of interest to us is to determine how the change of
the zero--range parameters affects the responses calculated within a
given theory. This will inform us about the necessity of considering
or not in detail the role of these parameters. This aspect is analyzed
by considering $V^{\rm II}_{\rm res}$ with parameters $g_0$ and 
$g_0^\prime$ fixed, as in the case of the J\"ulich--Stony Brook 
interaction, to reproduce some low energy properties in the lead
region (see details in the next section). It is worth to point out
that $V^{\rm II}_{\rm res}$ is precisely the interaction used in 
practice in much of the calculations mentioned above and that is why we 
want to use it for this analysis.

Our analysis focuses on the transverse response functions in the
quasi--elastic peak. We will not consider the longitudinal ones
because they are strongly influenced by the spin independent pieces of
the interaction (in particular, the $f_0$ and $f_0^\prime$ channels)
and these are difficult to fix at low energy because of the role
played by the scalar mesons not usually taken into account. 

\section{RESULTS OF THE CALCULATIONS}

The investigation of the various questions we are interested in has been 
carried out by comparing different calculations of the transverse (e,e')
responses in $^{40}$Ca for three different momentum transfer ($q=300$,
410 and 550~MeV/$c$).

\newpage

\vspace*{1cm}

\begin{center}
\setlength{\unitlength}{1mm}
\begin{picture}(100,180)(35,0)
\put(0,0){\ewxy{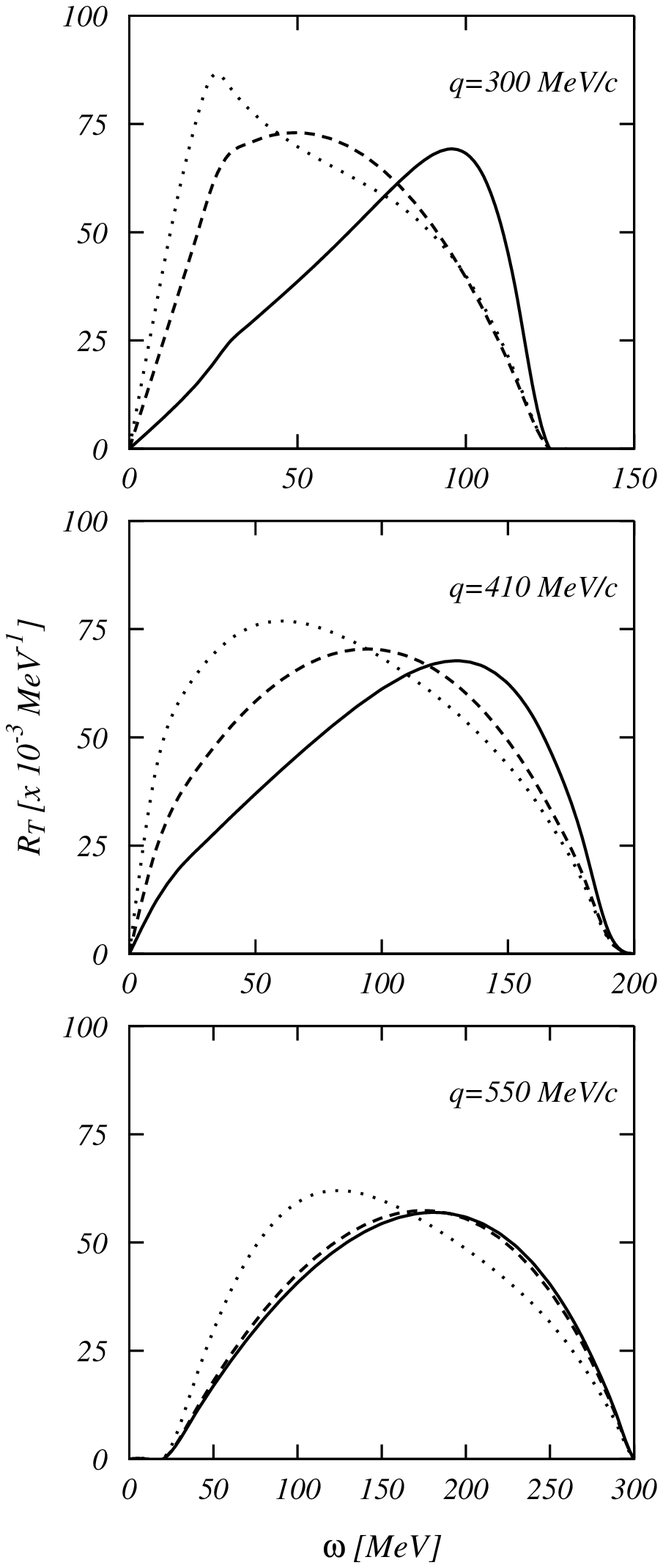}{180mm}}
\end{picture} 
\end{center} 

\vspace*{-1.2cm}
\noindent
{\small {\sc Fig.~1}. Transverse nuclear responses for $^{40}$Ca, 
calculated for the three momentum transfers we have considered in this
work. Dotted lines correspond to an RPA calculation with 
$V^{\rm I}_{\rm res}$, while solid curves represent the RA results for
$V^{\rm II}_{\rm res}$. In both cases the values $g_0=0.57$ and 
$g_0^\prime=0.717$ (with $C_0=386.04$~MeV~fm$^3$) have been used. 
Dashed curves give the free FG responses. In all the calculations a 
value of $k_{\rm F}=235$~MeV/$c$ has been used.}
\vspace{1cm}

\newpage

First we have study the effects produced when an effective
interaction, which has been determined in a given effective theory 
(e.g. RPA), is used to calculate (e,e') transverse responses in a 
different framework (e.g. RA).

By considering the parameterization of Ref.~\cite{sp80} (that is 
$g_0=0.57$, $g_0^\prime=0.717$ and $C_0=386.04$~MeV~fm$^3$), we have 
carried out two different calculations, the results of which are shown
Fig.~1. Therein, solid curves correspond to the calculations performed
in the FG approach within the RA and with the interaction 
$V^{\rm II}_{\rm res}$ in Eq.~(\ref{int-II}). On the other hand, 
dotted lines have been obtained within the RPA, also for the FG. The 
model used in this case is the one developed in Ref.~\cite{ba96}, 
which, contrary to what happens for the RA approach, includes 
explicitly the exchange terms in the RPA expansion. In this case we 
have used the interaction $V^{\rm I}_{\rm res}$ in Eq.~(\ref{int-I}) 
and we have adopted the factor $r=0.4$, which is consistent with the 
parameterization used. Also in Fig.~1, we have plotted the free FG 
responses for comparison (dashed curves). 

The first comment one can draw from these results is that the use of 
the interaction, as it was fixed at low energy, leads to transverse 
responses which are quite different from those obtained in the RA
(with $V^{\rm II}_{\rm res}$) calculation, though the differences 
reduce with increasing momentum transfer. As we can see, the results 
obtained in the RPA are peaked at lower energies and this is a clear 
evidence of a more attractive residual interaction. It is 
straightforward to check this point because the central piece of the 
$V^{\rm I}_{\rm res}$ is attractive, while the contrary happens for
$V^{\rm II}_{\rm res}$, at least for $q\leq 2k_{\rm F}$. On the other 
hand it is interesting to note how the RA results are more similar to 
the free response as long as $q$ increases, while the same does not 
occurs for the RPA responses. 

Obviously, the reason for the discrepancies between both calculations 
can be ascribed to the two basic ingredients of the effective theories
used in each case: the exchange terms, which are included in the RPA 
calculations but not in the RA ones, and the reduction factor $r$ 
modifying the $\rho-$exchange potential.

Before going deeper in this question, it is worth to comment on the
nuclear wave functions used in the calculations discussed above. As in
any FG type calculation, plane--waves have been considered here to 
describe the single--particle states. The fact that the interaction was 
fixed in a framework which considered microscopic RPA wave functions,
based on Woods--Saxon single--particle states, is an obvious 
inconsistency. Despite that, it has been shown~\cite{am94,am96}
that, in this energy region, the details concerning the nuclear wave 
functions are not extremely important and, at least to some extent, the 
shell--model response can be reasonably described with the FG model, 
provided an adequate value of the Fermi momentum, $k_{\rm F}$, is
used. In the present work, where we study the response in $^{40}$Ca,
we have taken $k_{\rm F}=235$~MeV/$c$ which gives a good agreement 
between FG and finite nuclei calculations~\cite{am94}.

We come back to investigate the reasons for the large discrepancy
between the RPA and RA calculations presented above. To do that we have 
done two new calculations: RA with $V^{\rm I}_{\rm res}$ and RPA with 
$V^{\rm II}_{\rm res}$. These calculations have been compared with the
two previous ones by means of the two following quantities:
\begin{equation}
\label{relexc}
\displaystyle
\gamma_{\rm exc}^r(q,\omega) \, = \,
\displaystyle
\frac{R_T^{{\rm RPA}(r)}(q,\omega) \, - \, 
      R_T^{{\rm RA}(r)}(q,\omega) } 
{R_T^{{\rm RA}(r)}(q,\omega)}
\end{equation}
and
\begin{equation}
\label{relfac}
\displaystyle
\gamma_r^{\rm mod}(q,\omega) \, = \,
\displaystyle
\frac{R_T^{{\rm mod}(r=1.0)}(q,\omega) \, - \, 
      R_T^{{\rm mod}(r=0.4)}(q,\omega) } 
{R_T^{{\rm mod}(r=0.4)}(q,\omega) } \, . 
\end{equation}

The first one gives us information about the effect of the consideration
of the exchange terms in the calculation. The corresponding results have
been plotted in Fig.~2 (left panels). The first aspect to be noted is 
that the exchange terms produce effects considerably larger for 
$V^{\rm I}_{\rm res}$ (solid lines) than for $V^{\rm II}_{\rm res}$ 
(dashed curves). These effects reduce with increasing momentum
transfer and they are rather small for $V^{\rm II}_{\rm res}$ above
$q=410$~MeV/$c$. 

On the other hand, the effect of the reduction factor $r$ in the 
$\rho$--exchange potential is measured with the parameter $\gamma_r$. 
The values of this parameter for the two effective models considered, 
these are RPA and RA, are shown in Fig.~2 (right panels), with solid
and dashed curves respectively. It is apparent that the effects of 
considering the $r$ factor are much larger than those due to the 
exchange. In general they are more important for the RA calculations 
than for the RPA ones, and reduce the higher $q$ is. 

\newpage

\vspace*{1cm}

\begin{center}
\setlength{\unitlength}{1mm}
\begin{picture}(100,180)(30,0)
\put(0,0){\ewxy{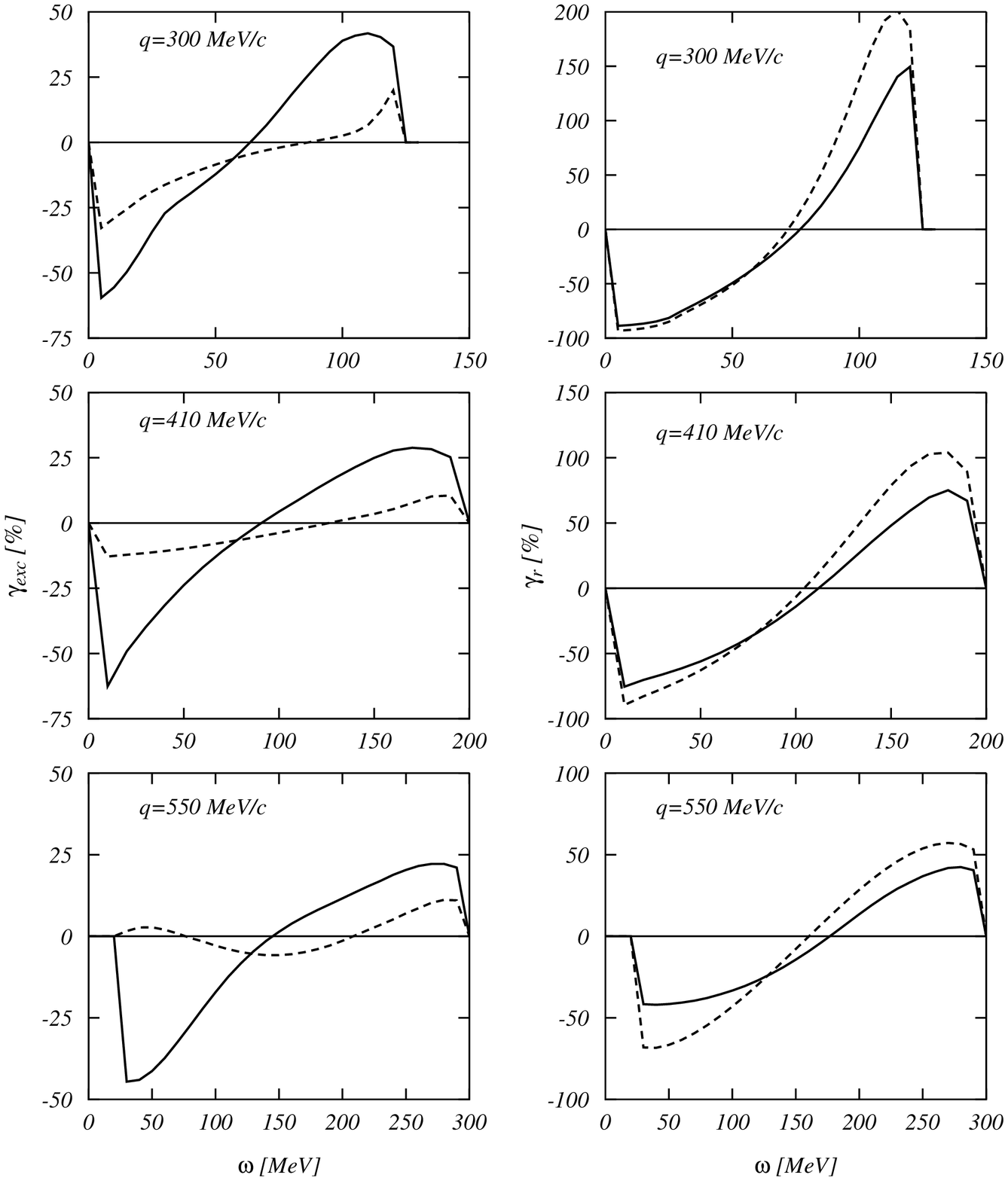}{180mm}}
\end{picture} 
\end{center} 

\vspace*{-1.2cm}
\noindent
{\small {\sc Fig.~2}. Left panels: $\gamma_{\rm exc}$, in percentage, 
as defined in Eq.~(\ref{relexc}). Dashed (solid) curves give the results 
obtained for $r=1.0 (0.4)$. Right panels: $\gamma_r$, in \%, calculated 
as in Eq.~(\ref{relfac}), for RPA (solid curves) and RA (dashed curves).}
\vspace{1cm}

\newpage

The first conclusion to be noted is that when using a given
interaction is mandatory to take care of the effective theory where 
its parameterization was fixed. The change of the framework produces 
results which could not be under control.

The open question in this respect is how different becomes the 
responses calculated within different effective theories but with an
interaction fixed consistently with the theory. This is the second
aspect we investigate. To do that we have considered the 
$V^{\rm II}_{\rm res}$ and have determined the parameters $g_0$ and 
$g_0^\prime$ of the Landau--Migdal piece in such a way that the 
energies and B-values of the two $1^+$ states in $^{208}$Pb at 5.85 
and 7.30~MeV are reproduced. This has been done both in RPA and RA. 
The reason for choosing these two states lies in their respective 
isoscalar and isovector character, what makes them particularly 
adequate to permit the determination of both parameters almost 
independently. The values obtained in this procedure are shown in 
Table~I. It is remarkable the small value of $g_0$ needed for the RA
calculation. A similar result is found when a pure zero--range 
Landau--Migdal interaction is adjusted, with the same criterion, in 
RPA type calculations (see Refs.~\cite{co90,hi92}). This points out
the importance of the exchange, at least at low energy.

\vspace{.5cm}
\noindent
{\small {\sc TABLE I.} Values of the Landau--Migdal parameters $g_0$ and 
$g_0^\prime$ obtained in the procedure of fixing the effective interaction 
$V^{\rm II}_{\rm res}$ (see text). The values quoted ``RPA'' (``RA'') 
correspond to calculations performed with (without) the consideration
of the exchange terms.}
\begin{center}
\begin{tabular}{cccc}
\hline\hline
Effective theory && $g_0$ & $g_0^\prime$ \\ \hline
RPA && $~0.470$ & $~0.760$ \\
 RA && $~0.038$ & $~0.717$ \\\hline\hline
\end{tabular}
\end{center}

\vspace{1cm}

With the interaction fixed in this way we have evaluated the
transverse responses for the three momentum transfer we are
considering throughout this work. The results are shown in Fig.~3
where dotted (solid) curves correspond to the RPA (RA) calculations.
Dashed lines represent the free FG responses. As we can see, the
differences between the results obtained with the two effective
theories are now much smaller than in Fig.~1.

\newpage

\vspace*{1cm}

\begin{center}
\setlength{\unitlength}{1mm}
\begin{picture}(100,180)(35,0)
\put(0,0){\ewxy{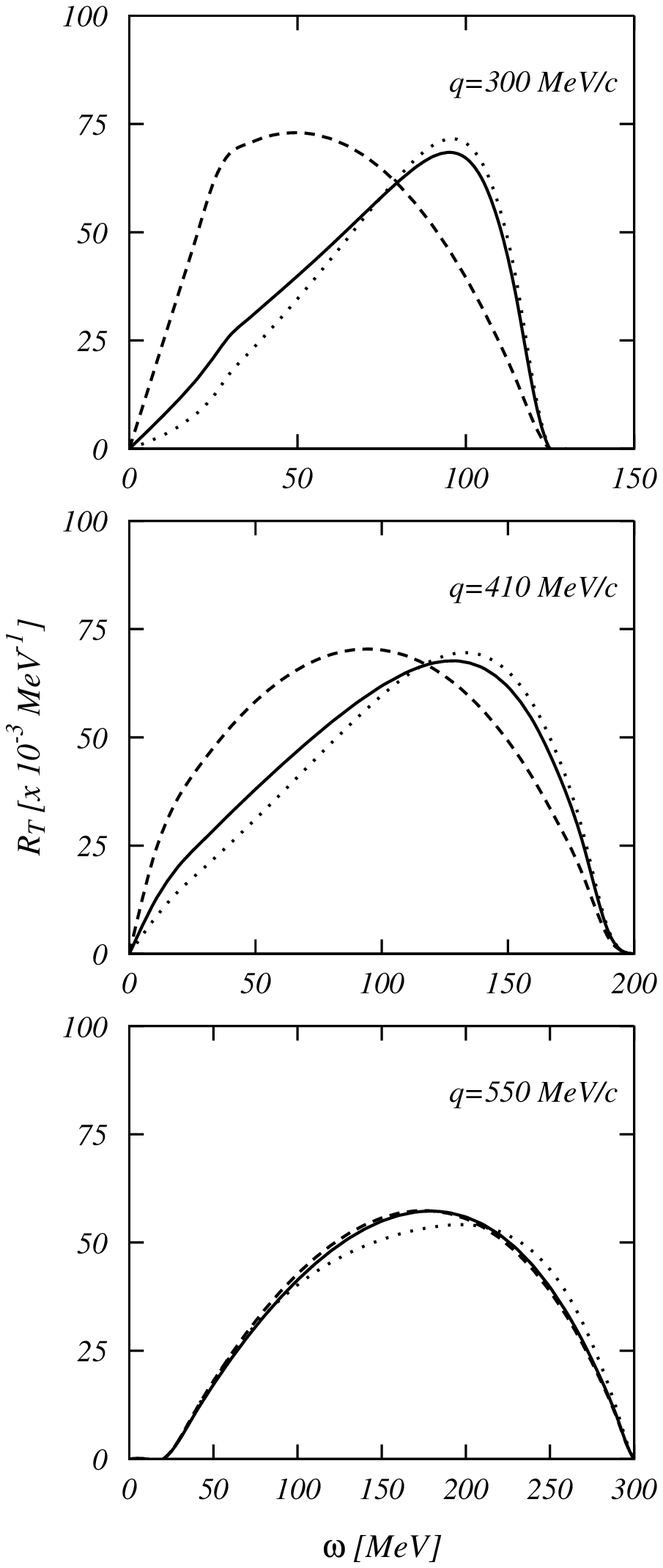}{180mm}}
\end{picture} 
\end{center} 

\vspace*{-1.2cm}
\noindent
{\small {\sc Fig.~3}. Transverse nuclear responses for $^{40}$Ca,
calculated with the $V^{\rm II}_{\rm res}$ interaction. Dotted lines 
correspond to an RPA calculation while solid curves represent the RA 
results. The values of $g_0$ and $g_0^\prime$ in Table~I have been 
used. Dashed curves give the free FG responses. In all the 
calculations a value of $k_{\rm F}=235$~MeV/$c$ has been used.}

\vspace{1cm}

\newpage

Two points deserve a comment. First, it is clear that the large 
differences observed between the RPA calculation here discussed and
that shown in Fig.~1 are mainly due to the presence of the reduction 
factor $r=0.4$ in the $V^{\rm I}_{\rm res}$ interaction. Second,
the similitude of the results obtained with the two calculations done
now with $V^{\rm II}_{\rm res}$, shows up the relevance of the link 
between effective theories and interactions.

The last aspect we want to analyze is how the responses calculated in a
given approach change when the zero--range parameters are modified. In
other words, we want to determine what is the role of these
parameters. How $g_0^\prime$ affects the responses is a point which
has been investigated with a certain detail in different previous
works (see e.g., Ref.~\cite{ba96}) and then we focus here in $g_0$.
Its influence can be seen in Fig.~4, where we compare the responses
plotted in Fig.~3 (solid curves), with those obtained by changing the 
$g_0$ parameter in order to use values considered by different
authors. Dashed--dotted curves correspond to $g_0=0$. Dashed lines 
represent the responses obtained with $g_0=0.70$ (0.57) for the RPA 
(RA) calculation. The values of $g_0^\prime$ have not been changed. 
The first point to be noted is the insensibility of the RA responses 
to the changes in $g_0$. As we can see, strong changes in $g_0$ 
produce almost no effect on the RA result. This can be easily 
understood because in the ring series the $g_0$ contribution is 
weighted with the magnetic moment $\mu_s^2$ while the $g_0^\prime$ 
piece appears with $\mu_v^2$. That means, the $g_0$ contribution is
$\mu_s^2/\mu_v^2 \, \approx \, 1/28$ of the $g_0^\prime$
contribution. The situation is different in the RPA case, where the 
$g_0$ contribution is as important as the $g_0^\prime$ one because 
of the presence of the exchange terms (see Ref.~\cite{ba96}). This 
makes that some of the RA calculations performed by other authors can 
be considered as ``consistent'' in practice, of course despite the 
fact that these parametrizations are unable to reproduce low energy 
properties. For example, in Ref.~\cite{gi97}, the parameterization of 
the J\"ulich--Stony Brook interaction was considered and this 
coincides with one of those used here ($g_0=0.57$ and 
$g_0^\prime=0.717$).

The results obtained in this work open a series of questions which we
consider worth for nuclear calculations in this energy region. In the
following we enumerate and comment them:

\newpage

\vspace*{.5cm}
\begin{center}
\setlength{\unitlength}{1mm}
\begin{picture}(100,170)(24,0)
\put(0,0){\ewxy{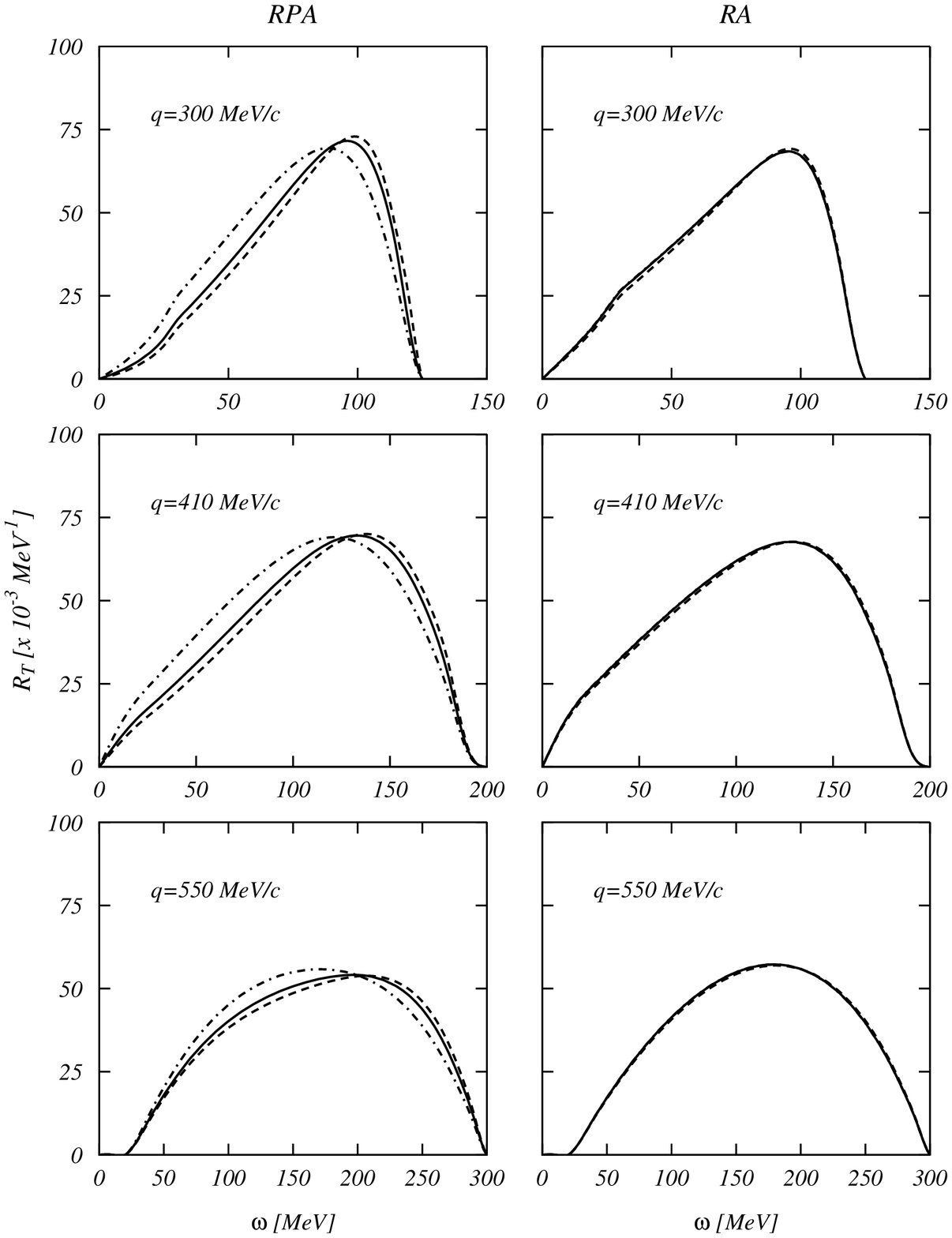}{170mm}}
\end{picture} 
\end{center} 

\vspace*{-1.5cm}
\noindent
{\small {\sc Fig.~4}. $R_T$ responses calculated in the RPA (left
panels) and RA (right panels). Solid curves correspond to
the parametrizations of Table~I. Dashed--dotted curves have been 
obtained with $g_0=0$, while the dashed ones correspond to 
$g_0=0.70$ (0.57) for the RPA (RA) calculation,
with the same values of $g_0^\prime$ as for the solid curves.}

\newpage

\begin{enumerate}

\item It has been shown that the strength of the tensor piece of 
$V^{\rm I}_{\rm res}$ is too strong to describe low energy properties 
(see, e.g., Ref.~\cite{co90}) and different mechanisms have been 
proposed to cure this problem (core--polarization effects~\cite{na85},
two--particles two--holes excitations~\cite{dr86}, {\it in--medium} 
scaling law~\cite{br91}, etc.) The role of the tensor part of the 
interaction in the quasi-elastic peak should be investigated in order 
to establish the effective force to be used.

\item The presence of the exchange terms increase the sensitivity of 
the responses to the details of the interaction. How important can be
the interference between these terms and other physical mechanisms
basic in this energy region (such as, e.g., meson--exchange currents,
final state interactions, short--range correlations, etc.) is a matter
of relevance in order to fully understand the nuclear response. The
analysis of the possible differences between RA and RPA with respect 
to these effects is of special interest in view of the fact that RA 
calculations are the most usual in the quasi--elastic peak.

\item The procedure of fixing the interaction is basic in order to deal
with the possibility of having an unique framework to calculate the 
nuclear response at any momentum transfer and excitation energy. The 
problem of developing such ``unified'' model is still unsolved, but
the cross analysis of low energy nuclear properties and quasi--elastic
peak responses could give valuable hints.

\end{enumerate} 

\section{CONCLUSIONS}

In this work we have analyzed the role of the effective interaction in
the quasi--elastic peak region by comparing the results obtained with
different effective theories and forces previously fixed in order to 
give a reasonable description of several low energy nuclear properties.

Some conclusions can be drawn after our analysis. First, it has been 
found that the interaction plays a role that, similarly to what
happens at low excitation energy, cannot be neglected. The particular 
point to be noted is the necessity of using effective interactions 
which have been fixed within an effective theory.

Second, the procedure we have followed to perform the calculations, 
that is to determine the interaction at low energy before calculating
at the quasi--elastic peak, seems to be adequate to look for an
``unique'' framework to calculate the nuclear response in different
energy and momentum regimes. 

The role of the tensor piece of the interaction must be
investigated. At low energy is a basic ingredient of the nuclear
structure calculations. Thus it is important to disentangle its
contribution in other excitation energy regions. Additionally, it 
seems encouraging to analyze the problem by including other physical 
mechanisms (meson--exchange currents, short--range correlations, 
final state interactions, etc.) which are known to be important in 
the description of the nuclear response and which depend on the 
interaction. 

\acknowledgements

Discussions with G. Co' are kindly acknowledged.
This work has been supported in part by the DGES (Spain) under 
contract PB95-1204 and by the Junta de Andaluc\'{\i}a (Spain).


\begin{references}

\bibitem[*]{AAAuth}Fellow of the Consejo Nacional
de Investigaciones Cient\'{\i}ficas y T\'ecnicas, CONICET.

\bibitem{al84}
W.M. Alberico, M. Ericson and A. Molinari, Ann. Phys. (N.Y.)
{\bf 154}, 356 (1984);
W.M. Alberico, A. De Pace, A. Drago and A. Molinari, Rivista del
Nuovo Cimento {\bf 14}, 1 (1991).

\bibitem{ry88}
J. Ryckebusch,	M. Warroquier, K. Heyde, J. Moreau and D. Ryckbosch,
Nucl. Phys. {\bf A476}, 237 (1988).

\bibitem {de91}
A. De Pace and M. Viviani, Phys. Rev. C {\bf 48}, 2931 (1993).

\bibitem {bu91}
M. Buballa, S. Dro\.zd\.z, S. Krewald and J. Speth, Ann. Phys. (N.Y.)
{\bf 208}, 346 (1991);
M. Buballa, S. Dro\.zd\.z, S. Krewald and A. Szczurek, Phys. Rev. C
{\bf 44}, 810 (1991);
S. Jeschonnek, A. Szczurek, G. Co' and S. Krewald, Nucl. Phys. 
{\bf A570}, 599 (1994).

\bibitem {ba95}
E. Bauer, Nucl. Phys. {\bf A589}, 669 (1995).

\bibitem {sl95}
V. Van der Sluys, J. Ryckebusch and M. Warroquier,
Phys. Rev. C {\bf 51}, 2664 (1995).

\bibitem {ba96}
E. Bauer, A. Ramos and A. Polls, Phys. Rev. {\bf C54}, 2959 (1996).

\bibitem {de96}
M.B. Barbaro, A. De Pace, T.W. Donnelly and A. Molinari,
Nucl. Phys. {\bf A596}, 553 (1996); {\it ibid.} {\bf A598}, 503 (1996).

\bibitem {gi97}
A. Gil, J. Nieves and E. Oset, Nucl. Phys. {\bf A} (1997) (in press); 
A. Gil, Ph. D. Thesis, Universitat de Val\`encia, 1996.

\bibitem {sp80}
J. Speth, V. Klemt, J. Wambach and G. E. Brown,
Nucl. Phys. {\bf A343}, 382 (1980).

\bibitem {am94}
J.E. Amaro, A.M. Lallena and G. Co', Int. J. Mod. Phys.
{\bf E3}, 735 (1994).

\bibitem {am96}
J.E. Amaro {\it et al.}, Nucl. Phys. {\bf A602}, 263 (1996).

\bibitem {co90}
G. Co' and A.M. Lallena, Nucl. Phys. {\bf A510}, 139 (1990).

\bibitem {hi92}
N.M. Hintz, A.M. Lallena and A. Sethi, Pys. Rev. C {\bf 45}, 1098 
(1992).

\bibitem {na85}
K. Nakayama, Phys. Lett. {\bf B165}, 239 (1985).

\bibitem{dr86}
S. Dro\.zd\.z, J.L Ta\'{\i}n and J. Wambach, Phys. Rev. C {\bf 34},
345 (1986).

\bibitem {br91}
G.E Brown and M. Rho, Phys. Rev. Lett. {\bf 66}, 2720 (1991).

\end{references}
\end{document}